\documentclass[sigconf]{acmart}

\copyrightyear{2023} 
\acmYear{2023} 
\setcopyright{acmlicensed}\acmConference[KDD '23]{Proceedings of the 29th ACM SIGKDD Conference on Knowledge Discovery and Data Mining}{August 6--10, 2023}{Long Beach, CA, USA}
\acmBooktitle{Proceedings of the 29th ACM SIGKDD Conference on Knowledge Discovery and Data Mining (KDD '23), August 6--10, 2023, Long Beach, CA, USA}
\acmPrice{15.00}
\acmDOI{10.1145/3580305.3599256}
\acmISBN{979-8-4007-0103-0/23/08}

\usepackage[ruled,linesnumbered]{algorithm2e} 
\usepackage{graphicx} 
\usepackage{balance}
\usepackage{subfig} 
\usepackage{amsfonts}
\usepackage{amsmath}
\usepackage{url}
\usepackage{mathrsfs}
\usepackage{subfig} 
\usepackage{multirow}
\usepackage{booktabs} 
\usepackage{makecell}
\usepackage{threeparttable}
\usepackage[normalem]{ulem}

\newcommand\eqcite[1]{Equation~\eqref{#1}}
\usepackage{graphicx}
\usepackage{multirow}

\begin{document}


\title{All in One: Multi-Task Prompting for Graph Neural Networks}

\author{Xiangguo Sun}
\affiliation{%
  \institution{Department of Systems Engineering and Engineering Management, 
and Shun Hing Institute of Advanced Engineering, The Chinese University of Hong Kong}
  \country{}
  }
\email{xiangguosun@cuhk.edu.hk}

\author{Hong Cheng}
\affiliation{
  \institution{Department of Systems Engineering and Engineering Management, and Shun Hing Institute of Advanced Engineering, 
The Chinese University of Hong Kong}
  \country{}
  }
\email{hcheng@se.cuhk.edu.hk}

\author{Jia Li}
\affiliation{%
  \institution{Data Science and Analytics Thrust, The Hong Kong University of Science and Technology (Guangzhou)}
  \country{}
  }
\email{jialee@ust.hk}

\author{Bo Liu}
\affiliation{%
  \institution{School of Computer Science and Engineering, Southeast University}
  \institution{Purple Mountain Laboratories}
  \country{}
  }
\email{bliu@seu.edu.cn}

\author{Jihong Guan}
\affiliation{%
  \institution{Department of Computer Science and Technology,
Tongji University}
  \country{}
  }
\email{jhguan@tongji.edu.cn}


\begin{abstract}
Recently, ``pre-training and fine-tuning'' has been adopted as a standard workflow for many graph tasks since it can take general graph knowledge to relieve the lack of graph annotations from each application. However, graph tasks with node level, edge level, and graph level are far diversified, making the pre-training pretext often incompatible with these multiple tasks. This gap may even cause a ``negative transfer'' to the specific application, leading to poor results. Inspired by the prompt learning in natural language processing (NLP), which has presented significant effectiveness in leveraging prior knowledge for various NLP tasks, we study the prompting topic for graphs with the motivation of filling the gap between pre-trained models and various graph tasks. In this paper, we propose a novel multi-task prompting method for graph models. Specifically, we first unify the format of graph prompts and language prompts with the prompt token, token structure, and inserting pattern. In this way, the prompting idea from NLP can be seamlessly introduced to the graph area. Then, to further narrow the gap between various graph tasks and state-of-the-art pre-training strategies, we further study the task space of various graph applications and reformulate downstream problems to the graph-level task. Afterward, we introduce meta-learning to efficiently learn a better initialization for the multi-task prompt of graphs so that our prompting framework can be more reliable and general for different tasks. We conduct extensive experiments, results from which demonstrate the superiority of our method.
\end{abstract}




\maketitle

\section{Introduction}
Graph neural networks (GNNs) have been widely applied to various applications such as social computing 
\cite{sun2022self, chen2022brainnet}
,  anomaly detection 
\cite{BWGNN,sun2022structure}
, and network analysis \cite{chen2020multi}. Beyond exploring various exquisite GNN structures, recent years have witnessed a new research trend on how to train a graph model for dedicated problems.  

\begin{figure}[t]
    \centering
    \includegraphics[width =0.35\textwidth]{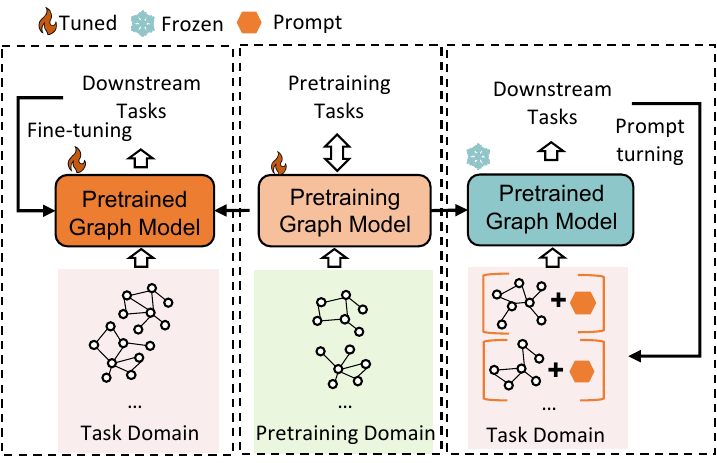}
    \caption{Fine-tuning, Pre-training, and Prompting.
    }
    \label{fig:prompt_tuning}
\end{figure}

Traditional supervised learning methods on graphs heavily rely on graph labels, which are not always sufficient in the real world. Another shortcoming is the over-fitting problem when the testing data is out-of-distribution \cite{shen2021towards}. To solve these challenges, many studies turn to ``pre-training and fine-tuning'' \cite{jin2020self}, which means pre-training a graph model with easily accessible data, and then transferring the graph knowledge to a new domain or task via tuning the last layer of the pre-trained model. Although much progress has been achieved on pre-training strategies \cite{hao2021multi}, there still exists a huge gap between these pretexts and multiple downstream tasks. For example, a typical pretext for the pre-training graph is binary edge prediction. Usually, this pre-training strategy makes connected nodes closer in a latent representation space. However, many downstream tasks are not limited to edge-level tasks but also include node-level tasks (e.g., node multi-class classification) or graph-level tasks (e.g., graph classification). If we transfer the above pre-trained model to multi-class node classification, it may require us to carefully search the results in higher dimensional parameter space for the additional classes of node labels. 
This tuning may even break down (a.k.a negative transfer \cite{wang2021afec}) when connected nodes have different labels. 
Tuning this pre-trained model to graph-level tasks is neither efficient because we have to pay huge efforts to learn an appropriate function mapping node embedding to the whole graph representation.

A promising solution to the above problems is to extend ``pre-training and fine-tuning'' to ``pre-training, prompting, and fine-tuning''. Prompt learning is a very attractive idea derived from natural language processing (NLP) and has shown notable effectiveness in generalizing pre-trained language models to a wide range of language applications \cite{min2021recent}. Specifically, a language prompt refers to a piece of text appended to the rear of an input text. For example, a sentiment task like ``\textit{KDD2023 will witness many high-quality papers.} \underline{I feel so} [MASK]'' can be easily transferred to a word prediction task via a preset prompt (``\underline{I feel so} [MASK]''). It is highly expected that the language model may predict ``[MASK]'' as ``excited'' rather than ``upset'' without further optimizing parameters for the new sentiment task because this model has already been pre-trained via the pretext of masked words prediction and contains some useful knowledge to answer this question. By this means, some downstream objectives can be naturally aligned with the pre-training target. Inspired by the success of the language prompt, we hope to introduce the same idea to graphs. As shown in Figure \ref{fig:prompt_tuning}, prompt tuning in the graph domain is to seek some light-weighted prompt, keep the pre-training model frozen, and use the prompt to reformulate downstream tasks in line with the pre-training task. In this way, the pre-trained model can be easily applied to downstream applications with highly efficient fine-tuning or even without any fine-tuning. This is particularly useful when the downstream task is a few-shot setting.

\begin{figure}[t]
    \centering
    \includegraphics[width =0.49\textwidth]{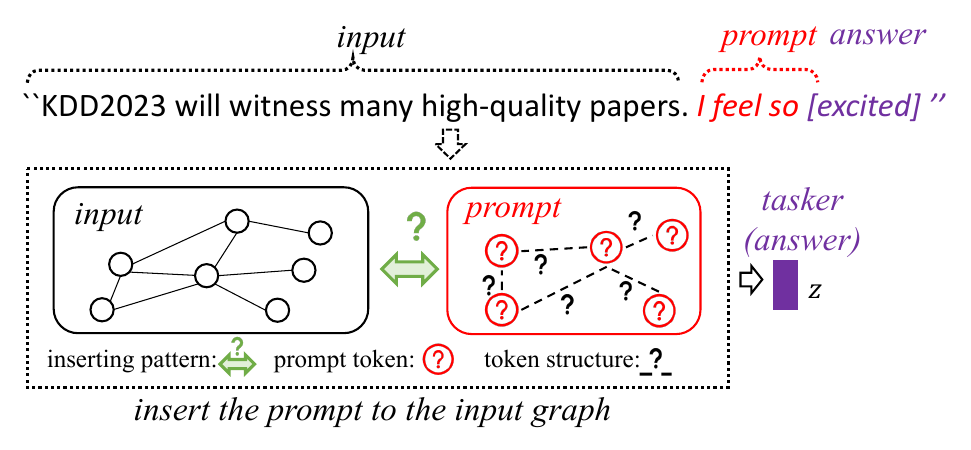}
    \caption{Our graph prompt inspired by the language prompt.}
    \label{fig:analogy}
\end{figure}

However, designing the graph prompt is more intractable than language prompts. First, classic language prompts are usually some preset phrases or learnable vectors attached at the end of input texts. As shown in Figure \ref{fig:analogy}, we only need to consider the content for the language prompt, whereas the graph prompt not only requires the prompt ``content'' but also needs to know how to organize these prompt tokens and how to insert the prompt into the original graph, both of which are undefined problems.
 
Second, there is a huge difficulty in reconciling downstream problems to the pre-training task. In the NLP area, we usually pre-train a language model via masked prediction and then transfer it to various applications like question answering \cite{rogers2020getting}, sentiment classification \cite{ling2022vision}. The underlying support \cite{qin2021exploring} is that these language tasks usually share a large overlapping task sub-space, making a masked language task easily transferred to other applications. However, how much does the same 
observation exist (if truly exists) in graph learning? 
It is crucial but difficult to decide on an appropriate pre-training task and reformulate downstream tasks to improve the capability of model generalization. Currently, we only find very few works \cite{sun2022gppt} studying the graph prompt issue. However, it can only deal with a single-type task (e.g., node classification) using a specific pretext (e.g., edge prediction), which is far from addressing the multi-task setting with different-level tasks.

Last but not least, learning a reliable prompt usually needs huge manpower and is more sensitive to prompt initialization in the multi-task setting \cite{liu2021pre}. Although there are some works \cite{lester2021power, zhong2021factual} in the NLP area trying to initialize the prompt via hand-crafted content or some discrete features, these methods are task-bounded, which is not sufficient when we confront a new task. This problem may be even worse in our multi-task graph area since graph features vary a lot in different domains and tasks. 

\textit{\textbf{Presented work.}}
To further fill the gap between graph pre-training and downstream tasks, we introduce the prompt method from NLP to graphs under the multi-task background. Specifically, to address the first challenge, we propose to unify the format of the language prompt and graph prompt in one way so that we can smoothly transfer the prompt idea from NLP to graphs, then we design the graph prompt from prompt tokens, token structures, and prompt inserting patterns. To address the second challenge, we first study the task subspace in graphs and then propose to reformulate node-level and edge-level tasks to graph-level tasks by induced graphs from original graphs. To address the third challenge, we introduce the meta-learning technique over multiple tasks to learn better prompts. We carefully evaluate our method with other approaches and the experimental results extensively demonstrate our advantages. 

\textit{\textbf{Contributions:}}
\begin{itemize}
    \item We unify the prompt format in the language area and graph area, and further propose an effective graph prompt for multi-task settings (section \ref{subsec:prompt_graph}).
    \item We propose an effective way to reformulate node-level and edge-level tasks to graph-level tasks, which can further match many pre-training pretexts (section \ref{subsec:refor}). 
    \item We introduce the meta-learning technique to our graph prompting study so that we can learn a reliable prompt for improving the multi-task performance (section \ref{subsec:meta}).
    \item We carefully analyze why our method works (section \ref{subsec:why}) and confirm the effectiveness of our method via extensive experiments (section \ref{sec:eva}).
\end{itemize}

\section{Background}
\noindent \textbf{Graph Neural Networks.} Graph neural networks (GNNs) have presented powerful expressiveness in many graph-based applications 
\cite{sun2021multi, li2019predicting, hou2022graphmae, Cheng2023wiener}
. The nature of most GNNs is to capture the underlying message-passing patterns for graph representation. To this end, there are many effective neural network structures proposed such as graph attention network (GAT) \cite{velivckovic2018graph}, graph convolution network (GCN) \cite{welling2016semi}, Graph Transformer \cite{shi2020masked}. Recent works also consider how to make graph learning more adaptive when data annotation is insufficient or how to transfer the model to a new domain, which triggered many graph pre-training studies instead of traditional supervised learning.

\noindent \textbf{Graph Pre-training.} Graph pre-training \cite{jin2020self} aims to learn some general knowledge for the graph model with easily accessible information to reduce the annotation costs of new tasks. Some effective pre-training strategies include node-level comparison like GCA \cite{zhu2021graph}, edge-level pretext like edge prediction \cite{jin2020self}, and graph-level contrastive learning such as GraphCL \cite{you2020graph} and SimGRACE \cite{xia2022simgrace}. In particular, GraphCL minimizes the distance between a pair of graph-level representations for the same graph with different augmentations whereas SimGRACE tries to perturb the graph model parameter spaces and narrow down the gap between different perturbations for the same graph. These graph-level strategies perform more effectively in graph knowledge learning \cite{hu2020strategies} and are the default strategies of this paper.

\noindent \textbf{Prompt Learning \& Motivations.} 
Intuitively, the above graph-level pre-training strategies have some intrinsic similarities with the language-masked prediction task: aligning two graph views generated by node/edge/feature mask or other perturbations is very similar to predicting some vacant ``blanks'' on graphs. That inspires us to further consider: why can't we use a similar format prompt for graphs to improve the generalization of graph neural networks?
Instead of fine-tuning a pre-trained model with an adaptive task head, prompt learning aims to reformulate input data to fit the pretext \cite{gao2021making}. 
Many effective prompt methods are firstly proposed in the NLP area, including some hand-crafted prompts like GPT-3 \cite{brown2020language}, discrete prompts like \cite{autoprompt,gao2021making}, and trainable prompts in the continuous spaces like \cite{li2021prefix, liu2022p}. Despite significant results achieved, prompt-based methods have been rarely introduced in graph domains yet. We only find very few works like GPPT \cite{sun2022gppt}, trying to design prompts for graphs. Unfortunately, most of them are very limited and are far from sufficient to meet the multi-task demands.

\section{Multi-task Prompting on Graphs}
\subsection{Overview of Our Framework}\label{subsec:overview}

\textit{\textbf{Objective: }} In this paper, we aim to learn a prompt graph that can be inserted into the original graph, through which we wish to further bridge the gap between a graph pre-training strategy and multiple downstream tasks, and further relieve the difficulties of transferring prior knowledge to different domains. 

\noindent \textit{\textbf{Overview: }} To achieve our goal, we propose a novel multi-task prompting framework for graph models. First, we unify various graph tasks in the same format and reformulate these downstream tasks as graph-level tasks. Second, with the unified graph-level instances, we further narrow down the gap among multiple tasks by a novel prompt graph with learnable tokens, inner structures, and adaptive inserting patterns. Third, we build a meta-learning process to learn more adaptive graph prompts for multi-task settings. Next, we elaborate on the main components.

\subsection{Reformulating Downstream Tasks}\label{subsec:refor}

\subsubsection{Why Reformulate Downstream Tasks.} The success of the traditional ``pre-training and fine-tuning'' framework in the NLP area largely lies in the fact that the pre-training task and downstream tasks share some common intrinsic task subspace, making the pre-training knowledge transferable to other downstream tasks (Figure \ref{fig:hierarchy_2}). However, things are a little complicated in the graph domain since graph-related tasks are far from similar. As shown in Figure \ref{fig:hierarchy_1}, it is far-fetched to treat the edge-level task and the node-level task as the same one because node-level operations and edge-level operations are far more different \cite{sun2022gppt}. This gap limits the performance of pre-training models and might even cause negative transfer \cite{jin2020self}. The same problem also happens in our ``pre-training, prompting, and fine-tuning'' framework since we aim to learn a graph prompt for multiple tasks, which means we need to further narrow down the gap between these tasks by reformulating different graph tasks in a more general form.

\subsubsection{Why Reformulate to the Graph Level.} 
\begin{figure}[t]
\centering
\subfloat[NLP tasks]{
\label{fig:hierarchy_2}
\includegraphics[width=0.16\textwidth]{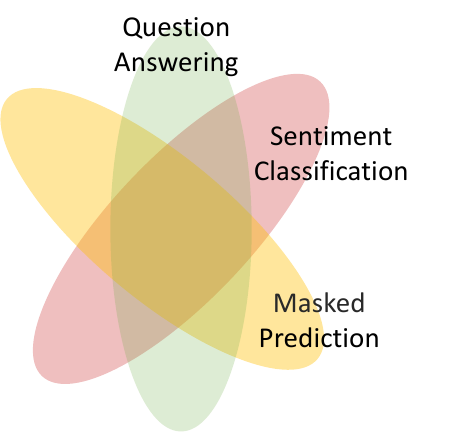}%
}
\subfloat[graph tasks]{
\label{fig:hierarchy_1}
\includegraphics[width=0.16\textwidth]{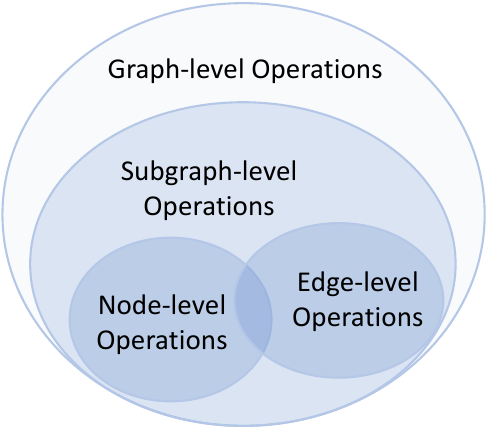}%
}
\caption{Task space in NLP and graph}
\label{fig:hierarchy}
\end{figure}

With the above motivation, we revisit the potential task space on the graph and find their hierarchical relation as shown in Figure \ref{fig:hierarchy_1}. Intuitively, many node-level operations such as ``changing node features'', ``delete/add a node'', or edge-level operations such as ``add/delete an edge'', can be treated as some basic operations at the graph level. For example, ``delete a subgraph'' can be treated as ``delete nodes and edges''. Compared with node-level and edge-level tasks, graph-level tasks are more general and contain the largest overlapping task sub-spaces for knowledge transfer, which has been adopted as the mainstream task in many graph pre-training models \cite{you2020graph, hu2020strategies, xia2022simgrace}. This observation further inspires us to reformulate downstream tasks to look like the graph-level task and then leverage our prompting model to match graph-level pre-training strategies.

\subsubsection{How to Reformulate Downstream Tasks.}\label{subsubsec:induced_graph}
\begin{figure}[h]
\centering
\subfloat[Induced graphs for nodes]{
\label{fig:ign}
\includegraphics[width=0.3\textwidth]{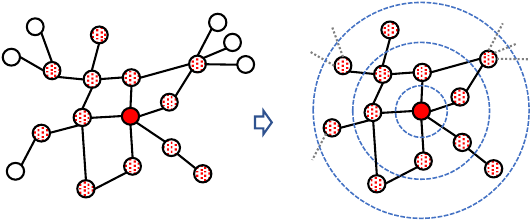}
}
\\
\subfloat[Induced graphs for edges]{
\label{fig:ige}
\includegraphics[width=0.3\textwidth]{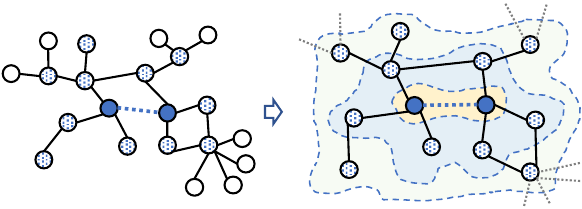}%
}
\caption{Induced graphs for nodes and edges}
\label{fig:ig}
\end{figure}

Specifically, we reformulate node-level and edge-level tasks to graph-level tasks by building induced graphs for nodes and edges, respectively. As shown in Figure \ref{fig:ign},
an induced graph for a target node means its local area in the network within $\tau$ distance, which is also known as its $\tau$-ego network. This subgraph preserves the node's local structure by neighboring node connections and its semantic context by neighboring node features, which is the main scope of most graph neural encoders. When we treat the target node's label as this induced graph label, we can easily translate the node classification problem into graph classification; Similarly, we present an induced graph for a pair of nodes in Figure \ref{fig:ige}. Here, the pair of nodes can be treated as a positive edge if there is an edge connecting them, or a negative edge if not. This subgraph can be easily built by extending this node pair to their $\tau$ distance neighbors. We can reformulate the edge-level task by assigning the graph label with the edge label of the target node pair. Note that for unweighted graphs, the $\tau$ distance is equal to $\tau$-hop length; for weighted graphs, the $\tau$ distance refers to the shortest path distance, where the induced graph can be easily found by many efficient algorithms \cite{akiba2015efficient,zhu2013efficient}.

\subsection{Prompt Graph Design}\label{subsec:prompt_graph}
\subsubsection{Prompting NLP and Graph in One Way} To seamlessly transfer the prompting idea from NLP to the graph domain, we propose to unify NLP Prompt and Graph Prompt in one perspective. Having compared the demand of NLP and graph area as shown in Figure \ref{fig:analogy}, we found that the prompt in NLP and graph areas should contain at least three components: 
\textit{(1) prompt token}, which contains the vectorized prompting information with the same size as the input word/node vector; \textit{(2) token structure}, which indicates the connection of different tokens. In the NLP area, prompt tokens (a.k.a prompt words) are preset as a linear relation like a sub-sentence or a phrase; whereas in the graph domain, the connections of different tokens are non-linear and far more complicated than NLP prompts; \textit{(3) inserting pattern}, which presents how to add the prompt to the input data. In the NLP area, the prompt is usually added in the front or the back end of the input sentences by default. However, in the graph area, there are no explicit positions like a sentence to joint graph prompt, making the graph prompting more difficult.

\subsubsection{Prompt Tokens}
Let a graph instance be $\mathcal{G}=(\mathcal{V},\mathcal{E})$ where $\mathcal{V}=\{v_1,v_2,\cdots,v_N\}$ is the node set containing $N$ nodes; each node has a feature vector denoted by $\mathbf{x}_i \in \mathbb{R}^{1\times d}$ for node $v_i$; $\mathcal{E}=\{(v_i,v_j)|v_i,v_j \in \mathcal{V}\}$ is the edge set where each edge connects a pair of nodes in $\mathcal{V}$. With the previous discussion, we here present our prompt graph as $\mathcal{G}_p=(\mathcal{P},\mathcal{S})$ where $\mathcal{P}=\{p_1,p_2,\cdots,p_{|\mathcal{P}|}\}$ denotes the set of prompt tokens and $|\mathcal{P}|$ is the number of tokens. Each token $p_i \in \mathcal{P}$ can be represented by a token vector $\mathbf{p}_i\in \mathbb{R}^{1\times d}$ with the same size of node features in the input graph; Note that in practice, we usually have $|\mathcal{P}|\ll N$ and $|\mathcal{P}|\ll d_h$ where $d_h$ is the size of the hidden layer in the pre-trained graph model. With these token vectors, the input graph can be reformulated by adding the $j$-th token to graph node $v_i$ (e.g., $\mathbf{\hat{x}}_i= \mathbf{x}_i +\mathbf{p}_j $). Then, we replace the input features with the prompted features and send them to the pre-trained model for further processing.

\subsubsection{Token Structures}
$\mathcal{S}=\{(p_i,p_j)|p_i,p_j \in \mathcal{P}\}$ is the token structure denoted by pair-wise relations among tokens. Unlike the NLP prompt, the token structure in the prompt graph is usually implicit. To solve this problem, we propose three methods to design the prompt token structures: (1) the first way is to learn tunable parameters:
$$\mathcal{A}=\overset{|\mathcal{P}|-1}{\underset{i=1 \atop j=i+1}{\cup}}\{a_{ij}\}$$

\noindent where $a_{ij}$ is a tunable parameter indicating how possible the token $p_i$ and the token $p_j$ should be connected; (2) the second way is to use the dot product of each prompt token pair and prune them according to the dot value. In this case, $(p_i,p_j)\in \mathcal{S}$ iff $\sigma(\mathbf{p}_i\cdot \mathbf{p}_j)<\delta$ where $\sigma(\cdot)$ is a sigmoid function and $\delta$ is a pre-defined threshold; (3) the third way is to treat the tokens as independent and then we have $\mathcal{S}=\emptyset$.

\subsubsection{Inserting Patterns}
Let $\psi$
be the inserting function that indicates how to add the prompt graph $\mathcal{G}_p$ to the input graph $\mathcal{G}$, then the manipulated graph can be denoted as $\mathcal{G}_m=\psi(\mathcal{G}, \mathcal{G}_p)$. We can define the inserting pattern as the dot product between prompt tokens and input graph nodes, and then use a tailored connection like
$\mathbf{\hat{x}}_i=\mathbf{x}_i+\sum_{k=1}^{|\mathcal{P}|} w_{ik}\mathbf{p}_k$ where $w_{ik}$ is a weighted value to prune unnecessary connections:
\begin{equation}
w_{ik}=\left\{
            \begin{array}{cl}
                 \sigma(\mathbf{p}_k\cdot \mathbf{x}_i^T), & \text{if } \sigma(\mathbf{p}_k\cdot \mathbf{x}_i^T)>\delta \\
                 0, & \text{otherwise}
            \end{array}
        \right.
\end{equation}
As an alternative and special case, we can also use a more simplified way to get $\mathbf{\hat{x}}_i=\mathbf{x}_i+\sum_{k=1}^{|\mathcal{P}|} \mathbf{p}_k$.

\subsection{Multi-task Prompting via Meta Learning}\label{subsec:meta}
\subsubsection{Constructing Meta Prompting Tasks}
Let $\tau_i$ be the $i$-th task with supporting data $\mathcal{D}^{s}_{\tau_i}$ and query data  $\mathcal{D}^{q}_{\tau_i}$; Specifically, for the graph classification task, $\mathcal{D}^{s}_{\tau_i}$ and $\mathcal{D}^{q}_{\tau_i}$ contain labeled graphs; for the node classification task, we generate an induced graph for each node as mentioned in section \ref{subsubsec:induced_graph}, align the graph label with the target node label, and treat this graph as a member in $\mathcal{D}^{s}_{\tau_i}$ or $\mathcal{D}^{q}_{\tau_i}$; for the edge classification task, we first generate edge induced graphs for training and testing and the edge label is up to its two endpoints.

\subsubsection{Applying Meta-learning to Graph Prompting}
Let $\theta$ be prompt parameters, $\pi^{*}$ be the fixed parameters of the pre-trained graph backbone, and $\phi$ be the tasker's parameters. We use $f_{\theta, \phi|\pi^{*}}$ to denote the pipeline with prompt graph ($\theta$), pre-trained model ($\pi^{*}$, fixed), and downstream tasker ($\phi$). Let $\mathcal{L}_{\mathcal{D}}(f)$ be the task loss with pipline $f$ on data $\mathcal{D}$. Then for each task $\tau_i$, the corresponding parameters can be updated as follows:
\begin{equation}
\begin{aligned}
\theta_i^k &=\theta_i^{k-1}-\alpha \nabla_{\theta_i^{k-1}} \mathcal{L}_{\mathcal{D}_{\tau_i}^{s}}\left(f_{\theta_i^{k-1}, \phi_i^{k-1}|\pi^{*}}\right) \\
\phi_i^k &=\phi_i^{k-1}-\alpha \nabla_{\phi_i^{k-1}}\mathcal{L}_{\mathcal{D}_{\tau_i}^{s}}\left(f_{\theta_i^{k-1}, \phi_i^{k-1}|\pi^{*}}\right) 
\end{aligned}
\end{equation}

\noindent where the initialization is set as: $\theta_i^0=\theta$, and $\phi_i^0=\phi$. The goal of this section is to learn effective initialization settings ($\theta, \phi$) for meta prompting tasks, which can be achieved by minimizing the meta loss on various tasks:
\begin{equation}
\theta^{*}, \phi^{*}=\underset{\theta, \phi}{\arg \min }\sum_{\tau_i \in \mathcal{T}} \mathcal{L}_{\mathcal{D}_{\tau_i}^{q}}\left(f_{\theta_i, \phi_i|\pi^{*}}\right) 
\end{equation}
where $\mathcal{T}$ is the task set. According to the chain rule, we use the second-order gradient to update $\theta$ (or $\phi$) based on query data:
\begin{equation}\label{equ:meta_update}
\begin{aligned}
\theta \!\leftarrow & \theta-\beta \cdot g_\theta^{second} \\
=& \theta-\beta \cdot \sum_{\tau_i \in \mathcal{T}} \nabla_\theta \mathcal{L}_{\mathcal{D}_{\tau_i}^{q}}\left(f_{\theta_i, \phi_i|\pi^{*}}\right) \\
=&\theta-\beta \cdot \sum_{\tau_i \in \mathcal{T}} \nabla_{\theta_i} \mathcal{L}_{\mathcal{D}_{\tau_i}^{q}}\left(f_{\theta_i, \phi_i|\pi^{*}}\right) \cdot \nabla_\theta\left(\theta_i\right) \\
=&\theta\!-\beta\! \cdot \!\!\!\!\sum_{\tau_i \in \mathcal{T}} \!\!\! \nabla_{\theta_i} \mathcal{L}_{\mathcal{D}_{\tau_i}^{q}}\!\!\left(f_{\theta_i, \phi_i|\pi^{*}}\right)\!\cdot \!\left(\mathbf{I}\!-\!\alpha \mathbf{H}_\theta\!\left(\mathcal{L}_{\mathcal{D}_{\tau_i}^{s}}\left(f_{\theta_i, \phi_i|\pi^{*}}\!\right)\!\right)\!\right)
\end{aligned}
\end{equation}

\noindent where $\mathbf{H}_\theta(\mathcal{L})$ is the Hessian matrix with $(\mathbf{H}_\theta(\mathcal{L}))_{ij}=\partial^2\mathcal{L}/\partial\theta_i\theta_j$; and $\phi$ can be updated in the same way. 

Kindly note that in the prompt learning area, the task head is also known as the answering function, which connects the prompt to the answers for downstream tasks to be reformulated. The answering function can be both tunable or hand-craft templates. In section \ref{subsec:why}, we also propose a very simple but effective hand-crafted prompt answering template without any tunable task head.

\subsubsection{Overall Learning Process}
To improve the learning stability, we organize these tasks as multi-task episodes where each episode contains batch tasks including node classification (``$n$'' for short), edge classification (``$\ell$'' for short), and graph classification (``$g$'' for short). Let $\mathcal{E}_i=(\mathcal{T}_{\mathcal{E}_i}, \mathcal{L}_{\mathcal{E}_i}, \mathcal{S}_{\mathcal{E}_i},\mathcal{Q}_{\mathcal{E}_i})$ be a multi-task episode. We define task batch  $\mathcal{T}_{\mathcal{E}_i}=\{\mathcal{T}^{(g)}_{\mathcal{E}_i},\mathcal{T}^{(n)}_{\mathcal{E}_i},\mathcal{T}^{(\ell)}_{\mathcal{E}_i}\}$ where each subset $\mathcal{T}^{(\triangleleft)}_{\mathcal{E}_i}=\{\tau_{\triangleleft1},\cdots,\tau_{\triangleleft t_{\triangleleft}}\}$; loss function sets $\mathcal{L}_{\mathcal{E}_i}=\{\mathcal{L}^{(g)},\mathcal{L}^{(n)},\mathcal{L}^{(\ell)}\}$, supporting data $\mathcal{S}_{\mathcal{E}_i}=\{\mathcal{S}_{\mathcal{E}_i}^{(g)},\mathcal{S}_{\mathcal{E}_i}^{(n)},\mathcal{S}_{\mathcal{E}_i}^{(\ell)}\}$ where each subset $\mathcal{S}_{\mathcal{E}_i}^{(\triangleleft)}=\{\mathcal{D}^s_{\tau_{\triangleleft1}},\cdots,\mathcal{D}^s_{\tau_{\triangleleft t_{\triangleleft}}}\}$, and query data $\mathcal{Q}_{\mathcal{E}_i}=\{\mathcal{Q}_{\mathcal{E}_i}^{(g)},\mathcal{Q}_{\mathcal{E}_i}^{(n)},\mathcal{Q}_{\mathcal{E}_i}^{(\ell)}\}$  where $\mathcal{S}_{\mathcal{E}_i}^{(\triangleleft)}=\{\mathcal{D}^q_{\tau_{\triangleleft1}},\cdots,\mathcal{D}^q_{\tau_{\triangleleft t_{\triangleleft}}}\}$. Then the multi-task prompting is presented in Algorithm \ref{alg:prompting}.  We treat each node/edge/graph class as a binary classification task so that they can share the same task head. 
Note that our method can also deal with other tasks beyond classification with only a few adaptations (see Appendix \ref{sec:app}).

\begin{algorithm}[t]
\KwIn{Overall pipeline $f_{\theta, \phi|\pi^{*}}$ with prompt parameter $\theta$, pre-trained model with frozen parameter $\pi^{*}$, and task head parameterized by $\phi$; Multi-task episodes $\mathcal{E}=\{\mathcal{E}_1,\cdots,\mathcal{E}_n\}$;
}
\KwOut{
Optimal pipeline $f_{\theta^{*}, \phi^{*}|\pi^{*}}$
}
Initialize $\theta$ and $\phi $\\
\While{not done}
{
\tcp{inner adaptation}
Sample $\mathcal{E}_i \in \mathcal{E}$ where $\mathcal{E}_i=(\mathcal{T}_{\mathcal{E}_i}, \mathcal{L}_{\mathcal{E}_i}, \mathcal{S}_{\mathcal{E}_i},\mathcal{Q}_{\mathcal{E}_i})$ \\
\For{$\tau_{\triangleleft t} \in \mathcal{T}_{\mathcal{E}_i},\triangleleft=g,n,\ell$}
{
$\theta_{\tau_{\triangleleft t}}, \phi_{\tau_{\triangleleft t}} \leftarrow \theta, \phi $\\

$\theta_{\tau_{\triangleleft t}} \leftarrow \theta_{\tau_{\triangleleft t}}-\alpha \nabla_{\theta_{\tau_{\triangleleft t}}} \mathcal{L}^{(\triangleleft)}_{\mathcal{D}_{\tau_{\triangleleft t}}^{s}}\left(f_{\theta_{\tau_{\triangleleft t}}, \phi_{\tau_{\triangleleft t}}|\pi^{*}}\right)$\\

$\phi_{\tau_{\triangleleft t}} \leftarrow \phi_{\tau_{\triangleleft t}}-\alpha \nabla_{\phi_{\tau_{\triangleleft t}}}\mathcal{L}^{(\triangleleft)}_{\mathcal{D}_{\tau_{\triangleleft t}}^{s}}\left(f_{\theta_{\tau_{\triangleleft t}}, \phi_{\tau_{\triangleleft t}}|\pi^{*}}\right) $\\

}
\tcp{outer meta update}

Update $\theta,\phi$ by \eqcite{equ:meta_update} on $\mathcal{Q}_{\mathcal{E}_i}=\{\mathcal{D}_{\tau_{\triangleleft t}}^{q}|\tau_{\triangleleft t} \in \mathcal{T}_{\mathcal{E}_i},\triangleleft=g,n,\ell\}$

} 
\Return{
$f_{\theta^{*}, \phi^{*}|\pi^{*}}$
}
\caption{Overall Learning Process}
\label{alg:prompting}
\end{algorithm}

\subsection{Why It Works?}\label{subsec:why}

\subsubsection{{Connection to Existing Work}}\label{subsubsec:con}
A prior study on graph prompt is proposed by \cite{sun2022gppt}, namely GPPT. They use edge prediction as a pre-training pretext and reformulate node classification to the pretext by designing labeled tokens added to the original graph. The compound graph will be sent to the pre-trained model again to predict the link connecting each node to the label tokens. Their work somehow is a special case of our method when our prompt graph only contains isolated tokens, each of which corresponds to a node category. However, there are at least three notable differences: (1) GPPT is not flexible to manipulate original graphs; (2) GPPT is only applicable for node classification; and (3) GPPT only supports edge prediction task as the pretext but is not compatible with more advanced graph-level pre-training strategies such as GraphCL \cite{you2020graph}, UGRAPHEMB \cite{bai2019unsupervised}, SimGRACE \cite{xia2022simgrace} etc. We further discuss these issues w.r.t.\ flexibility, efficiency, and compatibility as below.

\subsubsection{{Flexibility}}\label{subsec:flex}
The nature of prompting is to manipulate the input data to match the pretext. Therefore, the flexibility of data operations is the bottleneck of prompting performance. Let $g$ be any graph-level transformation such as ``changing node features'', ``adding or removing edges/subgraphs'' etc., and $\varphi^{*}$ be the frozen pre-trained graph model. For any graph $\mathcal{G}$ with adjacency matrix $\mathbf{A}$ and node feature matrix $\mathbf{X}$, Fang et al. \cite{fang2022prompt} have proved that we can always learn an appropriate prompt token $p^*$ making the following equation stand:
\begin{equation}\label{equ:error_bound_naive}
    \varphi^*\left(\mathbf{A}, \mathbf{X}+p^*\right)=\varphi^*({g}(\mathbf{A}, \mathbf{X}))+O_{p\varphi}
\end{equation}
This means we can learn an appropriate token applied to the original graph to imitate any graph manipulation. Here {\small $O_{p\varphi}$} denotes the error bound between the manipulated graph and the prompting graph w.r.t. their representations from the pre-trained graph model. This error bound is related to some non-linear layers of the model (\textit{unchangeable}) and the quality of the learned prompt (\textit{changeable}), which is promising to be further narrowed down by a more advanced prompt scheme. In this paper, we extend the standalone token to a prompt graph that has multiple prompt tokens with learnable inner structures. Unlike the indiscriminate inserting in \eqcite{equ:error_bound_naive} (``{\footnotesize $\mathbf{X}+p^*$}'' means the prompt token should be added to every node of the original graph), the inserting pattern of our proposed prompt graph is highly customized. Let $\psi(\mathcal{G}, \mathcal{G}_p)$ denote the inserting pattern defined in section \ref{subsec:prompt_graph}; $\mathcal{G}$ is the original graph, and $\mathcal{G}_p$ is the prompt graph, then we can learn an optimal prompt graph $\mathcal{G}_p^*$ to extend \eqcite{equ:error_bound_naive} as follows:
\begin{equation}\label{equ:error_bound_new}
    \varphi^*\left(\psi(\mathcal{G}, \mathcal{G}_p^*)\right)=\varphi^*(\mathbf{g}(\mathbf{A}, \mathbf{X}))+O^{*}_{p\varphi}
\end{equation}
By efficient tuning, the new error bound {\small $O^{*}_{p\varphi}$} can be further reduced. In section \ref{subsec:further}, we empirically demonstrate that {\small $O^{*}_{p\varphi}$} can be significantly smaller than {\small $O_{p\varphi}$} via efficient training. That means our method supports more flexible transformations on graphs to match various pre-training strategies.

\subsubsection{{Efficiency}} 
Assume an input graph has $N$ nodes and $M$ edges and the prompt graph has $n$ tokens with $m$ edges. Let the graph model contain $L$ layers and the maximum dimension of one layer be $d$. The parameter complexity of the prompt graph is only $O(nd)$. In contrast, some typical graph models (e.g., GAT \cite{velivckovic2018graph}) usually contain $O(LKd^2+LKd)$ parameters to generate node embedding and additional $O(Kd)$ parameters to obtain the whole graph representation ($K$ is the multi-head number). The parameters may be even larger in other graph neural networks (e.g., graph transformer \cite{yun2019graph}). In our prompt learning framework, we only need to tune the prompt with the pre-trained graph model frozen, making the training process converge faster than traditional transfer tuning.

For the time complexity, a typical graph model (e.g., GCN \cite{welling2016semi}) usually needs $O(LNd^2\!+\!LMd\!+\!Nd)$ time to generate node embedding via message passing and then obtain the whole graph representation (e.g., $O(Nd)$ for summation pooling). By inserting the prompt into the original graph, the total time is 
$O(L(n\!+\!N)d^2\!+\!L(m\!+\!M)d\!+\! (n\!+\!N)d)$. 
Compared with the original time, the additional time cost is only $O(Lnd^2\!+\!Lmd\!+\!nd)$ where $n\ll d, n\ll N, m\ll M$.

Besides the efficient parameter and time cost, our work is also memory friendly. Taking node classification as an example, the memory cost of a graph model largely includes parameters, graph features, and graph structure information. As previously discussed, our method only needs to cache the prompt parameters, which are far smaller than the original graph model. For the graph features and structures, traditional methods usually need to feed the whole graph into a graph model, which needs huge memory to cache these contents. However, we only need to feed an induced graph to the model for each node, the size of which is usually far smaller than the original graph. Notice that in many real-world applications, we are often interested in only a few parts of the total nodes, which means our method can stop timely if there is no more node to be predicted and we do not need to propagate messages on the whole graph either. This is particularly helpful for large-scale data.

\subsubsection{{Compatibility}}\label{subsub:com}
Unlike GPPT, which can only use binary edge prediction as a pretext, and is only applicable for node classification as downstream tasks, our framework can support node-level, edge-level, and graph-level downstream tasks, and adopt various graph-level pretexts with only a few steps of tuning. Besides, when transferring the model to different tasks, traditional approaches usually need to additionally tune a task head. In contrast, our method focuses on the input data manipulation and it relies less on the downstream tasks. This means we have a larger tolerance for the task head. For example, in section \ref{subsec:trans}, we study the transferability from other domains or tasks but we only adapt our prompt, leaving the source task head unchanged. We can even select some specific pretext and customize the details of our prompt without any tuned task head. Here we present a case that does not need to tune a task head and we evaluate its feasibility in section \ref{subsec:abla}.

\begin{center}
\fcolorbox{black!0}{gray!10}{\parbox{.42\textwidth}{
\textit{Prompt without Task Head Tuning:}

\textbf{Pretext:} GraphCL \cite{you2020graph}, a graph contrastive learning task that tries to maximize the agreement between a pair of views from the same graph.

\textbf{Downstream Tasks:} node/edge/graph classification.

\textbf{Prompt Answer:} \textit{node classification.} Assume there are $k$ categories for the nodes. We design the prompt graph with $k$ sub-graphs (a.k.a sub-prompts) where each sub-graph has $n$ tokens. Each sub-graph corresponds to one node category. Then we can generate $k$ graph views for all input graphs. We classify the target node with label $\ell$ ($\ell=1,2,\cdots, k$) if the $\ell$-th graph view is closest to the induced graph. It is similar to edge/graph classification.
}}
\end{center}

Interestingly, by shrinking the prompt graph as isolate tokens aligned with node classes and replacing the induced graphs with the whole graph, our prompt format can degenerate to GPPT, which means we can also leverage edge-level pretext for node classification. Since this format is exactly the same as GPPT, we will not discuss it anymore. Instead, we directly compare GPPT on node classification with our method.


\section{Evaluation}\label{sec:eva}
In this section, we extensively evaluate our method with other approaches on node-level, edge-level, and graph-level tasks of graphs. In particular, we wish to answer the following research questions: \textbf{Q1}: How effective is our method under the few-shot learning background for multiple graph tasks? \textbf{Q2}: How adaptable is our method when transferred to other domains or tasks? \textbf{Q3}: How do the main components of our method impact the performance? \textbf{Q4}: How efficient is our model compared with traditional approaches? \textbf{Q5}: How powerful is our method when we manipulate graphs?

\subsection{Experimental Settings}

\subsubsection{Datasets}: We compare our methods with other approaches on five public datasets including Cora \cite{welling2016semi}, CiteSeer \cite{welling2016semi}, Reddit \cite{hamilton2017inductive}, Amazon \cite{shchur2018pitfalls}, and Pubmed \cite{welling2016semi}. Detailed statistics are presented in Table \ref{tab:data} where the last column refers to the number of node classes. To conduct edge-level and graph-level tasks, we sample edges and subgraphs from the original data where the label of an edge is decided by its two endpoints and the subgraph label follows the majority of the subgraph nodes. For example, if nodes have 3 different classes, say $c_1, c_2, c_3$, then edge-level tasks contain at least 6 categories ($c_1,c_2,c_3, c_1c_2,c_1c_3,c_2c_3$). We also evaluate additional graph classification and link prediction on more specialized datasets where the graph label and the link label are inborn and not related to any node (see Appendix \ref{sec:app}).
\begin{table}[t]
\centering
\caption{Statistics of datasets.}
\label{tab:data}
\resizebox{0.35\textwidth}{!}{%
\begin{tabular}{@{}c|c|c|c|c@{}}
\toprule
Dataset & \#Nodes & \#Edges  & \#Features & \#Labels           \\ \midrule
Cora    & 2,708    & 5,429     & 1,433       & 7            \\
CiteSeer & 3,327   & 9,104    & 3,703        & 6  \\
Reddit  & 232,965  & 23,213,838 & 602        & 41           \\
Amazon  & 13,752   & 491,722   & 767        & 10            \\ 
Pubmed  & 19,717   & 88,648    & 500        & 3            \\ 
\bottomrule
\end{tabular}%
}
\end{table}

\subsubsection{Approaches}
Compared approaches are from three categories: \textbf{(1) Supervised methods}: these methods directly train a GNN model on a specific task and then directly infer the result. We here take three famous GNN models including GAT \cite{velivckovic2018graph}, GCN \cite{welling2016semi}, and Graph Transformer \cite{shi2020masked} (short as GT). These GNN models are also included as the backbones of our prompt methods. \textbf{(2) Pre-training with fine-tuning}: These methods first pre-train a GNN model in a self-supervised way such as GraphCL \cite{you2020graph} and SimGRACE \cite{xia2022simgrace}, then the pre-trained model will be fine-tuned for a new downstream task. \textbf{(3) Prompt methods}: With a pre-trained model frozen and a learnable prompt graph, our prompt method aims to change the input graph and reformulate the downstream task to fit the pre-training strategies.

\subsubsection{Implementations}
We set the number of graph neural layers as 2 with a hidden dimension of 100. To study the transferability across different graph data, we use SVD (Singular Value
    Decomposition) to reduce the initial features to 100 dimensions. The token number of our prompt graph is set as 10. We also discuss the impact of token numbers in section \ref{subsec:abla} where we change the token number from 1 to 20. We use the Adam optimizer for all approaches. The learning rate is set as 0.001 for most datasets. In the meta-learning stage, we split all the node-level, edge-level, and graph-level tasks randomly in 1:1 for meta-training and meta-testing. Reported results are averaged on all tested tasks. More implementation details are shown in Appendix \ref{sec:app}, in which we also analyze the performance on more datasets and more kinds of tasks such as regression, link prediction, and so on.

\begin{table*}[]
\centering
\caption{Node-level performance (\%) with 100-shot setting. IMP (\%): the average improvement of prompt over the rest.}
\label{tab:node_level}
\resizebox{0.9\textwidth}{!}{%
\begin{tabular}{@{}p{0.07\textwidth}<{\centering}c|p{0.02\textwidth}<{\centering}p{0.025\textwidth}<{\centering}p{0.035\textwidth}<{\centering}|p{0.025\textwidth}<{\centering}p{0.025\textwidth}<{\centering}p{0.035\textwidth}<{\centering}|p{0.025\textwidth}<{\centering}p{0.025\textwidth}<{\centering}p{0.035\textwidth}<{\centering}|p{0.025\textwidth}<{\centering}p{0.025\textwidth}<{\centering}p{0.035\textwidth}<{\centering}|p{0.025\textwidth}<{\centering}p{0.025\textwidth}<{\centering}p{0.035\textwidth}<{\centering}@{}}
\toprule
\multirow{2}{*}{\makecell[c]{Training\\ schemes}}     & \multirow{2}{*}{Methods} & \multicolumn{3}{c|}{Cora}       & \multicolumn{3}{c|}{CiteSeer}    & \multicolumn{3}{c|}{Reddit}     & \multicolumn{3}{c|}{Amazon} & \multicolumn{3}{c}{Pubmed}  \\
                                      &                          & Acc & F1 & AUC & Acc & F1 & AUC & Acc & F1 & AUC & Acc & F1 & AUC & Acc & F1 & AUC \\ \midrule
\multirow{3}{*}{supervised}           
& GAT                        & 74.45    & 73.21    & 82.97     & 83.00    & 83.20    & 89.33     & 55.64    & 62.03    & 65.38     & 79.00    & 73.42    & 97.81     & 75.00    & 77.56    & 79.72     \\
                                      & GCN                        & 77.55    & 77.45    & 83.71     & 88.00    & 81.79    & 94.79     & 54.38    & 52.47    & 56.82     & 95.36    & 93.99    & 96.23     & 53.64    & 66.67    & 69.89     \\
                                      & GT          & 74.25    & 75.21    & 82.04     & 86.33    & 85.62    & 90.13     & 61.50    & 61.38    & 65.56     & 85.50    & 86.01    & 93.01     & 51.50    & 67.34    & 71.91     
\\\midrule
\multirow{6}{*}{\makecell[c]{pre-train \\+\\  fine-tune}} 
& GraphCL+GAT                & 76.05    & 76.78    & 81.96     & 87.64    & 88.40    & 89.93     & 57.37    & 66.42    & 67.43     & 78.67    & 72.26    & 95.65     & 76.03    & 77.05    & 80.02     \\
                                      & GraphCL+GCN                & 78.75    & 79.13    & 84.90     & 87.49    & 89.36    & 90.25     & 55.00    & 65.52    & 74.65     & 96.00    & 95.92    & 98.33     & 69.37    & 70.00    & 74.74     \\
                                      & GraphCL+GT  & 73.80    & 74.12    & 82.77     & 88.50    & 88.92    & 91.25     & 63.50    & 66.06    & 68.04     & 94.39    & 93.62    & 96.97     & 75.00    & 78.45    & 75.05     \\
                                      & SimGRACE+GAT               & 76.85    & 77.48    & 83.37     & 90.50    & 91.00    & 91.56     & 56.59    & 65.47    & 67.77     & 84.50    & 84.73    & 89.69     & 72.50    & 68.21    & 81.97     \\
                                      & SimGRACE+GCN               & 77.20    & 76.39    & 83.13     & 83.50    & 84.21    & 93.22     & 58.00    & 55.81    & 56.93     & 95.00    & 94.50    & 98.03     & 77.50    & 75.71    & 87.53     \\
                                      & SimGRACE+GT & 77.40    & 78.11    & 82.95     & 87.50    & 87.05    & 91.85     & 66.00    & 69.95    & 70.03     & 79.00    & 73.42    & 97.58     & 70.50    & 73.30    & 74.22     
\\\midrule
\multirow{6}{*}{prompt}               
& GraphCL+GAT                & 76.50    & 77.26    & 82.99     & 88.00    & 90.52    & 91.82     & 57.84    & 67.02    & 75.33     & 80.01    & 75.62    & 97.96     & 77.50    & 78.26    & 83.02     \\
                                      & GraphCL+GCN                & 79.20    & 79.62    & 85.29     & 88.50    & 91.59    & 91.43     & 56.00    & 68.57    & 78.82     & 96.50    & 96.37    & 98.70     & 72.50    & 72.64    & 79.57     \\
                                      & GraphCL+GT  & 75.00    & 76.00    & 83.36     & 91.00    & 91.00    & 93.29     & 65.50    & 66.08    & 68.86     & 95.50    & 95.43    & 97.56     & 76.50    & 79.11    & 76.00     \\
                                      & SimGRACE+GAT               & 76.95    & 78.51    & 83.55     & 93.00    & 93.14    & 92.44     & 57.63    & 66.64    & 69.43     & 95.50    & 95.43    & 97.56     & 73.00    & 74.04    & 81.89     \\
                                      & SimGRACE+GCN               & 77.85    & 76.57    & 83.79     & 90.00    & 89.47    & 94.87     & 59.50    & 55.97    & 59.46     & 95.00    & 95.24    & 98.42     & 78.00    & 78.22    & 87.66     \\
                                      & SimGRACE+GT & 78.75    & 79.53    & 85.03     & 91.00    & 91.26    & 95.62     & 69.50    & 71.43    & 70.75     & 86.00    & 83.72    & 98.24     & 73.00    & 73.79    & 76.64  
\\ \midrule
\multicolumn{2}{c|}{IMP (\%)}                                        & 1.47     & 1.94     & 1.10      & 3.81     & 5.25     & 2.05      & 3.97     & 5.04     & 6.98      & 4.49&	5.84&	2.24  & 8.81     & 4.55     & 4.62      \\     
\midrule  
\multicolumn{2}{c|}{\makecell[c]{Reported Acc of GPPT (Label Ratio 50\%)}}        & 77.16    & --     & --      & 65.81     & --    & --     & 92.13     & --     & --      & 86.80  & -- & --  & 72.23    & --    & --      \\ \cline{1-2}
\multicolumn{2}{c|}{\makecell[c]{appr. Label Ratio of our 100-shot setting}}  & \multicolumn{3}{c|}{$\sim 25\% $}       & \multicolumn{3}{c|}{$\sim 18\% $}    & \multicolumn{3}{c|}{$\sim 1.7\% $}     & \multicolumn{3}{c|}{$\sim 7.3\% $} & \multicolumn{3}{c}{$\sim 1.5\% $}  \\
\bottomrule
\end{tabular}%
}
\end{table*}

\subsection{Multi-Task Performance with Few-shot Learning Settings (RQ1)}
We compared our prompt-based methods with other mainstream training schemes on node-level, edge-level, and graph-level tasks under the few-shot setting. We repeat the evaluation 5 times and report the average results in Table \ref{tab:node_level}, Table \ref{tab:edge_level} (Appendix \ref{sec:app}), and Table \ref{tab:graph_level} (Appendix \ref{sec:app}). From the results, we can observe that most supervised methods are very hard to achieve better performance compared with pre-train methods and prompt methods. This is because the empirical annotations required by supervised frameworks in the few-shot setting are very limited, leading to poor performance. In contrast, pre-training approaches contain more prior knowledge, making the graph model rely less on data labels. However, to achieve better results on a specific task, we usually need to carefully select an appropriate pre-training approach and carefully tune the model to match the target task, but this huge effort is not ensured to be applicable to other tasks. The gap between pre-training strategies and downstream tasks is still very large, making the graph model very hard to transfer knowledge on multi-task settings (we further discuss the transferability in section \ref{subsec:trans}.) Compared with pre-training approaches, our solutions further improve the compatibility of graph models. The reported improvements range from 1.10\% to 8.81\% on node-level tasks, 1.28\% to 12.26\% on edge-level tasks, and 0.14\% to 10.77\% on graph-level tasks. In particular, we also compared our node-level performance with the previously mentioned node-level prompt model GPPT in Table \ref{tab:node_level}. Kindly note that our experiment settings are totally different from GPPT. In GPPT, they study the few-shot problem by masking 30\% or 50\% data labels. However, in our paper, we propose a more challenging problem: how does the model perform if we further reduce the label data? So in our experiment, each class only has 100 labeled samples. This different setting makes our labeled ratio approximately only 25\% on Cora, 18\% on CiteSeer, 1.7\% on Reddit, 7.3\% on Amazon, and 1.5\% on Pubmed, which are far less than the reported GPPT (50\% labeled). 

\subsection{Transferability Analysis (RQ2)}\label{subsec:trans}
To evaluate the transferability, we compared our method with the hard transfer method and the fine-tuning method. Here the hard transfer method means we seek the source task model which has the same task head as the target task and then we directly conduct the model inference on the new task. The fine-tune method means we load the source task model and then tune the task head for the new task. We evaluate the transferability from two perspectives: (1) how effectively is the model transferred to different tasks within the same domain? and (2) how effectively is the model transferred to different domains?

\begin{table}[h]
\centering
\caption{Transferability (\%) on Amazon from different level tasks spaces. Source tasks: graph-level tasks and node-level tasks. Target task: edge-level tasks.}
\label{tab:trans_amazon}
\resizebox{0.4\textwidth}{!}{%
\begin{tabular}{@{}ll|lll@{}}
\toprule
Source task                  & Methods       & Accuracy & F1-score & AUC score \\ \midrule
\multirow{3}{*}{graph level} & hard & 51.50    & 65.96    & 40.34     \\
                             & fine-tune     & 62.50    & 70.59    & 53.91     \\
                             & prompt        & 70.50    & 71.22    & 74.02     \\ \midrule
\multirow{3}{*}{node level}  & hard & 40.50    & 11.85    & 29.48     \\
                             & fine-tune     & 46.00    & 54.24    & 37.26     \\
                             & prompt        & 59.50    & 68.73    & 55.90     \\ \bottomrule
\end{tabular}%
}
\end{table}
\begin{table}[h]
\centering
\caption{Transferability (\%) from different domains. Source domains: Amazon and PubMed. Target domain: Cora}
\label{tab:trans_domains}
\resizebox{0.45\textwidth}{!}{%
\begin{tabular}{@{}cc|ccc|ccc@{}}
\toprule
\multicolumn{2}{c|}{\makecell[c]{Source\\Domains}}    & \multicolumn{3}{c|}{Amazon}        & \multicolumn{3}{c}{PubMed}         \\ \midrule
Tasks                        &  & \makecell[c]{hard} & \makecell[c]{fine-tune} & prompt &\makecell[c]{hard} & \makecell[c]{fine-tune} & prompt  \\ \midrule
\multirow{3}{*}{\makecell[c]{node\\level}}  & Acc     & 26.9          & 64.14     & 65.07  & 55.62         & 57.93     & 62.07  \\
                             & F1      & 13.11         & 77.59     & 80.23  & 66.33         & 70.00     & 76.60  \\
                             & AUC     & 17.56         & 88.79     & 92.59  & 82.34         & 83.34     & 88.46  \\ \midrule
\multirow{3}{*}{\makecell[c]{edge\\level}}  & Acc     & 17.00         & 77.00     & 82.00  & 10.00         & 90.50     & 96.50  \\
                             & F1      & 10.51         & 81.58     & 84.62  & 2.17          & 89.73     & 91.80  \\
                             & AUC     & 4.26          & 94.27     & 96.19  & 6.15          & 93.89     & 94.70  \\ \midrule
\multirow{3}{*}{\makecell[c]{graph\\level}} & Acc     & 46.00         & 87.50     & 88.00  & 50.00         & 91.00     & 95.50  \\
                             & F1      & 62.76         & 89.11     & 88.12  & 10.00          & 93.90     & 95.60  \\
                             & AUC     & 54.23         & 86.33     & 94.99  & 90.85         & 91.47     & 98.47  \\ \bottomrule
\end{tabular}%
}
\end{table}

\subsubsection{Transferability to Different Level Tasks}
Here we pre-train the graph neural network on Amazon, then conduct the model on two source tasks (graph level and node level), and further evaluate the performance on the target task (edge level). For simplicity, both source tasks and the target task are built as binary classifications with $1:1$ positive and negative samples (we randomly select a class as the positive label and sample negatives from the rest). We report the results in Table \ref{tab:trans_amazon}, from which we have two observations: First, our prompt method significantly outperforms the other approaches and the prediction results make sense. In contrast, the problem of the hard transfer method is that the source model sometimes can not well decide on the target tasks because the target classes may be far away from the source classes. This may even cause negative transfer results (results that are lower than random guess). In most cases, the fine-tuning method can output meaningful results with a few steps of tuning but it can still encounter a negative transfer problem. Second, the graph-level task has better adaptability than the node-level task for the edge-level target, which is in line with our previous intuition presented in Figure \ref{fig:hierarchy} (section \ref{subsec:refor}). 

\subsubsection{Transferability to Different Domains} We also conduct the model on Amazon and PubMed as source domains, then load the model states from these source domains and report the performance on the target domain (Cora). Since different datasets have various input feature dimensions, we here use SVD to unify input features from all domains as 100 dimensions. Results are shown in Table \ref{tab:trans_domains}, from which we can find that the good transferability of our prompt also exists when we deal with different domains.

\begin{figure}[h]
    \centering
    \includegraphics[width =0.45\textwidth]{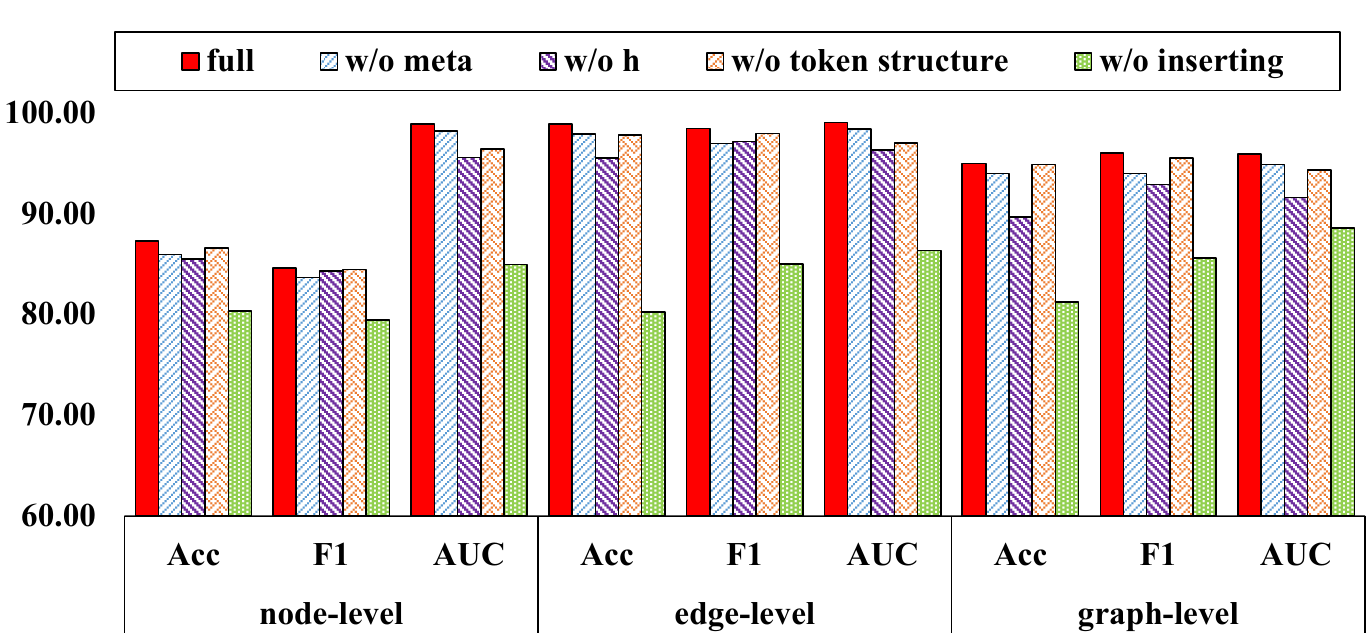}
    \caption{Effectiveness of main components}
    \label{fig:abl}
\end{figure} 

\subsection{Ablation Study (RQ3)}\label{subsec:abla}
In this section, we compare our complete framework with four variants: ``w/o meta'' is the prompt method without meta-learning step; ``w/o h'' is our method without task head tuning, which is previously introduced in section \ref{subsub:com}; ``w/o token structure'' is the prompt where all the tokens are treated as isolated without any inner connection; and ``w/o inserting'' is the prompt without any across links between prompt tokens and the input graphs. We report the performance in Figure \ref{fig:abl}, from which we can find the meta-learning and token structure all contribute significantly to the final results. In particular, the inserting pattern between a prompt graph and the input graph plays a very crucial role in the final performance. As previously discussed, the purpose of the prompt-based method is to relieve the difficulty of traditional ``pre-train, fine-tuning'' by filling the gap between the pre-training model and the task head. This means the prompt graph is proposed to further improve the fine-tuning performance. This is particularly important when we transfer the model across different tasks/domains, which proposes harder demand for the task head. As suggested in Figure \ref{fig:abl}, even when we totally remove the tunable task head, the ``w/o h'' variant can still perform very competitively, which suggests the powerful capability of bridging upstream and downstream tasks.

\subsection{Efficiency Analysis (RQ4)}
\begin{figure}[h]
    \centering
    \includegraphics[width =0.35\textwidth]{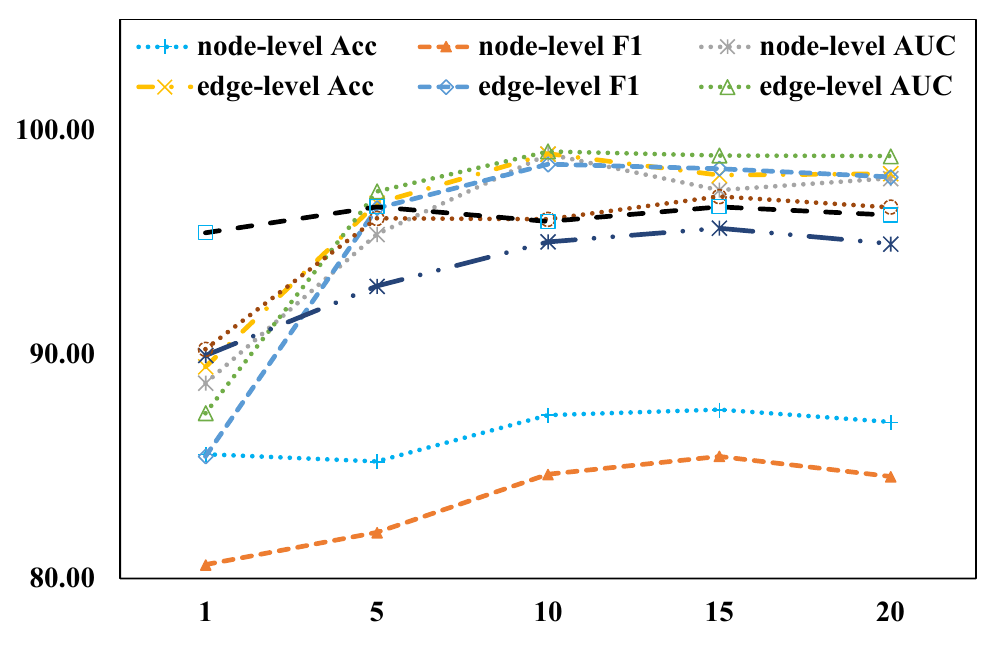}
    \caption{Impact of token numbers}
    \label{fig:token_num}
\end{figure}

\begin{figure}[h]
    \centering
    \includegraphics[width =0.33\textwidth]{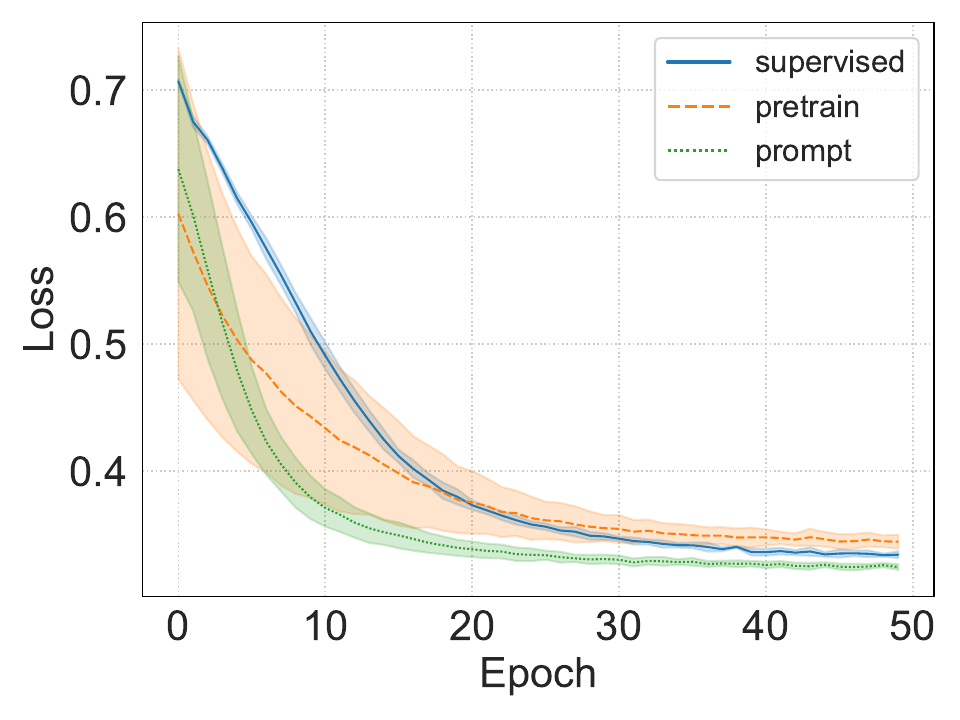}
    \caption{Training losses with epochs. Mean values and 65\% confidence intervals by 5 repeats with different seeds.}
    \label{fig:loss}
\end{figure}

Figure \ref{fig:token_num} presents the impact of increasing token number on the model performance, from which we can find that most tasks can reach satisfactory performance with very limited tokens, making the complexity of the prompt graph very small. The limited token numbers make our tunable parameter space far smaller than traditional methods, which can be seen in Table \ref{tab:num_par}. This means our method can be efficiently trained with a few steps of tuning. As shown in Figure \ref{fig:loss}, the prompt-based method converges faster than traditional pre-train and supervised methods, which further suggests the efficiency advantages of our method.
\begin{table}[h]
\centering 
\caption{Tunable parameters comparison. RED (\%): average reduction of the prompt method to others.}
\label{tab:num_par}
\resizebox{0.45\textwidth}{!}{%
\begin{tabular}{@{}c|ccccc|c@{}}
\toprule
Methods           & Cora & CiteSeer & Reddit & Amazon & Pubmed & RED (\%)\\ \midrule
GAT               & $\sim$ 155K  &  $\sim$ 382K      &  $\sim$ 75K     &  $\sim$ 88K     &  $\sim$ 61K & 95.4$\downarrow$  \\
GCN               &  $\sim$ 154K  &  $\sim$ 381K      &  $\sim$ 75K     &  $\sim$ 88K     &  $\sim$ 61K  & 95.4$\downarrow$ \\
GT &  $\sim$ 615K  &  $\sim$ 1.52M     &  $\sim$ 286K    &  $\sim$ 349K    &  $\sim$ 241K & 98.8$\downarrow$ \\
prompt            &  $\sim$ 7K    &  $\sim$ 19K       &  $\sim$ 3K      &  $\sim$ 4K      &  $\sim$ 3K &  --   \\ \bottomrule
\end{tabular}%
}
\end{table}

\subsection{Flexibility on Graph Transformation (RQ5)}\label{subsec:further}

\begin{figure}[h]
\centering 
\subfloat[pre-trained]{
\label{fig:init}
\includegraphics[width=0.24\textwidth]{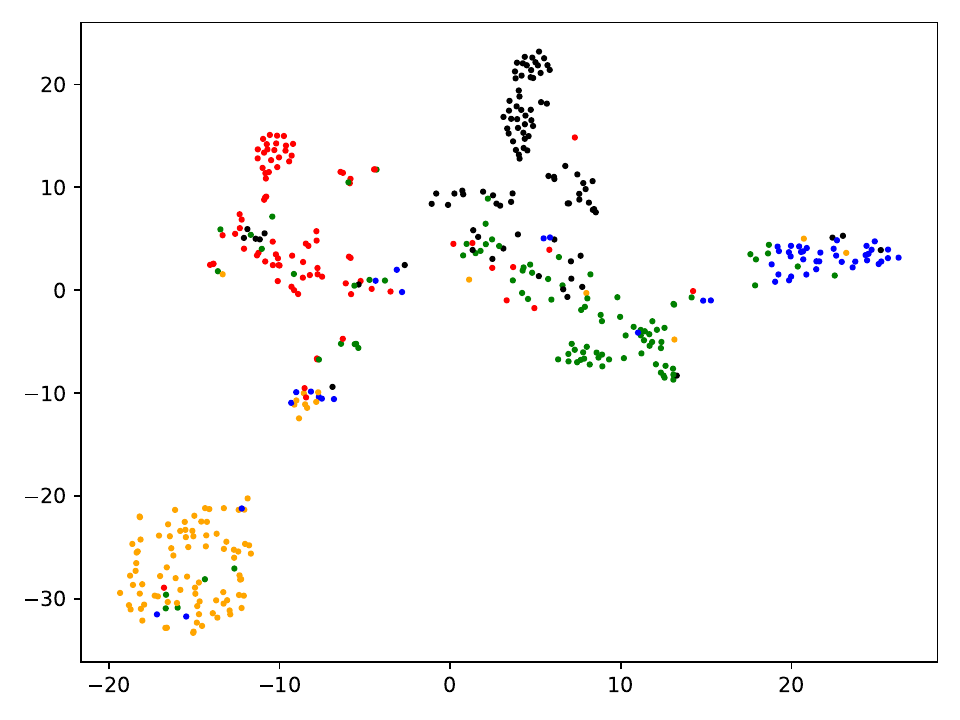}
}
\subfloat[prompt]{
\label{fig:prompt}
\includegraphics[width=0.24\textwidth]{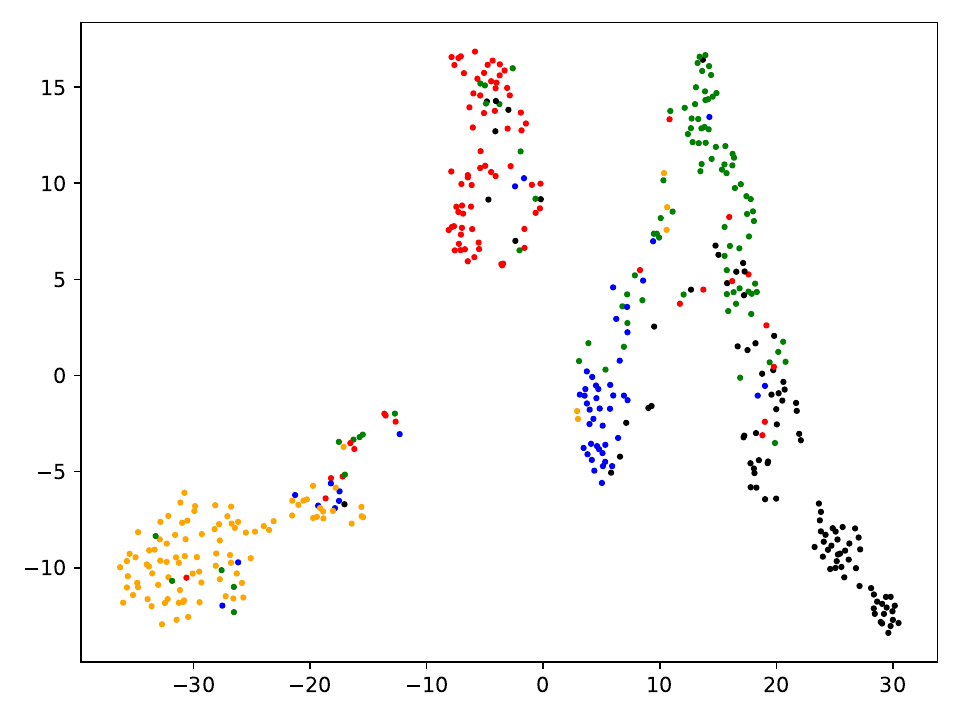}
}
\caption{Visualization of graph representations.}
\label{fig:vis}
\end{figure}

As discussed in section \ref{subsec:flex}, the flexibility of data transformation is the bottleneck of prompt-based methods. Here we manipulate several graphs by dropping nodes, dropping edges, and masking features, then we calculate the error bound mentioned in Equation \ref{equ:error_bound_naive} and \ref{equ:error_bound_new}. We compare the original error with the naive prompt mentioned in Equation \ref{equ:error_bound_naive}, and our prompt graph with 3, 5, and 10 tokens. As shown in Table \ref{tab:error}, our designed prompt graph significantly reduces the error between the original graph and the manipulated graph. This means our method is more powerful to stimulate various graph transformations and can further support significant improvement for downstream tasks. This capability can also be observed in the  graph visualization from two approaches. As shown in Figure \ref{fig:vis}, the graph representations from a pre-trained model present lower resolution to node classes compared with our prompted graph.

\section{Conclusion}
In this paper, we study the multi-task problem of graph prompts with few-shot settings. We propose a novel method to reformulate different-level tasks to unified ones and further design an effective prompt graph with a meta-learning technique. We extensively evaluate the performance of our method. Experiments demonstrate the effectiveness of our framework.

\begin{table}[h]
\centering 
\caption{Error bound discussed by section \ref{subsec:flex}
RED (\%): average reduction of each method to the original error.}
\label{tab:error}
\resizebox{0.45\textwidth}{!}{%
\begin{tabular}{@{}p{0.14\textwidth}<{\centering}c|ccc|c@{}}
\toprule
\makecell[c]{Prompt Solutions} & \makecell[c]{Token\\ Number} & \makecell[c]{Drop \\Nodes} & \makecell[c]{Drop \\Edges} & \makecell[c]{Mask \\Features} & RED (\%) \\ \midrule
\makecell[c]{Original Error \\(without prompt)} & 0        & 0.9917    & 2.6330    & 6.8209  &  -   \\\midrule
\makecell[c]{Naive Prompt \\(Equation \ref{equ:error_bound_naive})} & 1       & 0.8710    & 0.5241    & 2.0835  &  66.70$\downarrow$   \\\midrule
\multirow{3}{*}{\makecell[c]{Our Prompt Graph\\ (with token, structure,\\ and inserting patterns)}} & 3         & 0.0875    & 0.2337    & 0.6542  & 90.66$\downarrow$    \\
& 5         & 0.0685    & 0.1513    & 0.4372  & 93.71$\downarrow$    \\
& 10        & 0.0859    & 0.1144    & 0.2600  & 95.59$\downarrow$    \\ \bottomrule
\end{tabular}%
}
\end{table}

\begin{acks}

This research is supported in part by project \#MMT-p2-23 of the Shun Hing Institute of Advanced Engineering, The Chinese University of Hong Kong, by grants from the Research Grant Council of the Hong Kong Special Administrative Region, China (No. CUHK 14217622), NSFC (No. 61972087, No. 62206067, No. U1936205, No. 62172300, No. 62202336), Guangzhou-HKUST(GZ) Joint Funding Scheme (No. 2023A03J0673), National Key R\&D Program of China (No. 2022YFB3104300, No. 2021YFC3300300), the Fundamental Research Funds for the Central Universities (No. ZD-21-202101), and Open Research Projects of Zhejiang Lab (No. 2021KH0AB04). \textbf{The first author, Dr. Xiangguo Sun, in particular, wants to thank his parents for their kind support during his tough period.}

\end{acks}
\clearpage
\bibliographystyle{ACM-Reference-Format} 
\balance
\bibliography{ref.bib}

\clearpage

\appendix

\section{Appendix}\label{sec:app}
In this section, we supplement more experiments to evaluate the effectiveness of our framework further. 
The source code is publicly available at \textcolor{cyan}{\url{https://anonymous.4open.science/r/mpg}}

\begin{table}[t]
\centering
\caption{Statistics of Additional Datasets}
\label{tab:data_graph_more}
\resizebox{0.48\textwidth}{!}{%
\begin{tabular}{@{}p{0.11\textwidth}<{\centering}|p{0.057\textwidth}<{\centering}|p{0.057\textwidth}<{\centering}|p{0.061\textwidth}<{\centering}|p{0.047\textwidth}<{\centering}|c@{}}
\toprule
Dataset        & \#Nodes & \#Edges & \#Features & \#Labels & \#Graphs \\ \midrule
ENZYMES        & 19,580   & 74,564   & 21         & 6      & 600       \\
ProteinsFull & 43,471   & 162,088  & 32         & 2        & 1,113      \\\midrule
Movielens   &10,352 & 100,836  & 100         & -       & 1       \\
QM9            & 2,333,625 & 4,823,498 & 16         & -     & 129,433    \\\midrule 
PersonalityCafe   &  100,340   & 3,788,032 &   100      &   0  &  1     \\
Facebook  & 4,039 &  88,234    &  1,283  &    0  &    1  \\
\bottomrule
\end{tabular}%
}
\end{table}
\begin{table}[t]
\centering
\caption{Multi-class node classification (100-shots)}
\label{tab:app_multiclass}
\resizebox{0.48\textwidth}{!}{%
\begin{tabular}{@{}c|cc|cc@{}}
\toprule
\multirow{2}{*}{Methods} & \multicolumn{2}{c|}{Cora} & \multicolumn{2}{c}{CiteSeer} \\
     & Acc (\%) & Macro F1 (\%) & Acc (\%)& Macro F1  (\%)     \\ \midrule
Supervised    & 74.11  & 73.26        & 77.33          & 77.64 \\\midrule
Pre-train and Fine-tune & 77.97      & 77.63        & 79.67     & 79.83 \\ \midrule
Prompt     & 80.12      & 79.75        &  80.50   & 80.65    \\
Prompt w/o h    & 78.55          & 78.18            & 80.00     & 80.05  \\ \midrule
\makecell[c]{Reported Acc of GPPT\\ (Label Ratio 50\%)}  & 77.16      & -            & 65.81       & -   \\ 
\bottomrule
\end{tabular}%
}
\end{table}

\textbf{Additional Datasets} Besides the datasets mentioned in the main experiments of our paper, we here supplement more datasets in Table \ref{tab:data_graph_more} to further evaluate the effectiveness of our framework.
Specifically, ENZYMES and ProteinsFull are two molecule/protein datasets that are used in our additional graph-level classification tasks. 
Movielens and QM9 are used to evaluate the performance of our method on edge-level and graph-level regression, respectively. In particular, Movielens contains user's rating scores to the movies, each edge in which has a score value ranging from $0$ to $5$. 
QM9 is a molecule graph dataset where each graph has 19 regression targets, which are treated as graph-level multi-output regression. PersonalityCafe and Facebook datasets are used to test the performance of link prediction, both of which are social networks where edges denote the following/quoting relations. 

\textbf{Multi-label v.s. Multi-class Classification} In the main experiments, we treat the classification task as a multi-label problem. 
Here we present the experimental results under a multi-class setting. 
As reported in Table \ref{tab:app_multiclass}, our prompt-based method still outperforms the rest methods. 

\textbf{Additional Graph-level Classification} Here, we evaluate the graph-level classification performance where the graph label is not impacted by nodes' attributes. As shown in Table \ref{tab:app_graph_multiclass}, 
our method is more effective in the multi-class graph classification, especially in the few-shot setting.

\textbf{Edge/Graph-level Regression} Beyond classification tasks, our method can also support to improve graph models on regression tasks. Here, we evaluate the regression performance of both graph-level (QM9) and edge-level (MovieLens) datasets by MAE (mean absolute error) and MSE (mean squared error). We only feed 100-shot edge induced graphs for the model and the results are shown in Table \ref{tab:app_reg}, from which we can observe that our prompt-based methods outperform traditional approaches.

\textbf{Link Prediction} Beyond edge classification, link prediction is also a widely studied problem in the graph learning area. Here, the edges are split into three parts: (1) 80\% of the edges are for message passing only. (2) 10\% of the rest edges as the supervision training set. and (3) the rest edges as the testing set. For each edge in the training set and the testing set, we treat these edges as positive samples and sample non-adjacent nodes as negative samples. We generate the edge-induced graph for these node pairs according to the first part edges. The graph label is assigned as positive if the node pairs have a positive edge and vice versa. To further evaluate our method's potential in the extremely limited setting, we only sample 100 positive edges from the training set to train our model. In the testing stage, each positive edge has 100 negative edges. We evaluate the performance by MRR (mean reciprocal rank), and Hit Ratio@ 1, 5, 10. Results from Table \ref{tab:link} demonstrate that the performance of our prompt-based method still keeps the best in most cases.

\begin{table}[t]
\centering
\caption{Additional graph-level classification.}
\label{tab:app_graph_multiclass}
\resizebox{0.48\textwidth}{!}{%
\begin{tabular}{@{}c|cc|cc@{}}
\toprule
\multirow{2}{*}{Methods}  & \multicolumn{2}{c|}{ProteinsFull (100 shots)} & \multicolumn{2}{c}{ENZYMES (50 shots)} \\
     & Acc (\%)  & Macro F1 (\%)   & Acc (\%)& Macro F1 (\%)      \\ \midrule
Supervised    &  66.64 &  65.03   &  31.33  & 30.25 \\
Pre-train + Fine-tune & 66.50  &66.43    & 34.67   & 33.94  \\ 
Prompt     &70.50  & 70.17    & 35.00  & 34.92 \\
Prompt w/o h   & 68.50  & 68.50 &  36.67  &  34.05
\\ \bottomrule
\end{tabular}%
}
\end{table}

\begin{table}[t]
\centering
\caption{Graph/edge-level regression with few-shot settings.}
\label{tab:app_reg}
\resizebox{0.48\textwidth}{!}{%
\begin{tabular}{@{}c|cc|cc@{}}
\toprule
Tasks    & \multicolumn{2}{c|}{Graph Regression} & \multicolumn{2}{c}{Edge Regression} \\
Datasets & \multicolumn{2}{c|}{QM9 (100 shots)} & \multicolumn{2}{c}{MovieLens (100 shots)} \\
Methods     & MAE  & MSE  & MAE  & MSE     \\ \midrule
Supervised    & 0.3006  & 0.1327   &0.2285 & 0.0895   \\
Pre-train + Fine-tune & 0.1539  & 0.0351  &0.2171  &0.0774  \\ 
Prompt     &0.1384  & 0.0295   & 0.1949 &  0.0620   \\
Prompt w/o h   & 0.1424  &0.0341   & 0.2120    & 0.0744      
\\ \bottomrule
\end{tabular}%
}
\end{table}

\begin{table}[h]
\centering
\caption{Evaluation on link prediction (100-shot settings)}
\label{tab:link}
\resizebox{0.48\textwidth}{!}{%
\begin{tabular}{@{}c|cccc|cccc@{}}
\toprule
Datasets & \multicolumn{4}{c|}{PersonalityCafe} & \multicolumn{4}{c}{Facebook} \\
Methods     & MRR  & Hit@1 & Hit@5 & Hit@10& MRR  & Hit@1 & Hit@5 & Hit@10     \\ \midrule
Supervised  &0.18 & 0.04&0.24 &0.56  & 0.13&0.06 &0.17 &0.35 \\\midrule
\makecell[c]{Pre-train \\+ Fine-tune}  & 0.13&0.05 &0.12 &0.34  &0.10 &0.02 & 0.16& 0.33\\ \midrule
Prompt   &0.20 &0.07 &0.32 & 0.60 &0.19 &0.10 &0.23 &0.39  \\
Prompt w/o h   & 0.20&0.06 &0.30 & 0.50 &0.15 &0.09 &0.15 &0.33\\\midrule
\makecell[c]{Label Ratio}  & \multicolumn{4}{c|}{\makecell[c]{$\sim 0.003\%$ (training) \\ $\sim 80\%$(message passing)}}   & \multicolumn{4}{c}{\makecell[c]{$\sim 0.1\%$ (training) \\ $\sim 80\%$(message passing)}}  
\\ \bottomrule
\end{tabular}%
}
\end{table}




\begin{table*}[h]
\centering
\caption{Edge-level performance (\%) with 100-shot setting. IMP (\%): the average improvement of prompt over the rest.}
\label{tab:edge_level}
\resizebox{0.99\textwidth}{!}{%
\begin{tabular}{@{}p{0.07\textwidth}<{\centering}c|p{0.025\textwidth}<{\centering}p{0.025\textwidth}<{\centering}p{0.035\textwidth}<{\centering}|p{0.025\textwidth}<{\centering}p{0.025\textwidth}<{\centering}p{0.035\textwidth}<{\centering}|p{0.025\textwidth}<{\centering}p{0.025\textwidth}<{\centering}p{0.035\textwidth}<{\centering}|p{0.025\textwidth}<{\centering}p{0.025\textwidth}<{\centering}p{0.035\textwidth}<{\centering}|p{0.025\textwidth}<{\centering}p{0.025\textwidth}<{\centering}p{0.035\textwidth}<{\centering}@{}}
\toprule
\multirow{2}{*}{\makecell[c]{Training\\ schemes}}     & \multirow{2}{*}{Methods} & \multicolumn{3}{c|}{Cora}       & \multicolumn{3}{c|}{CiteSeer}    & \multicolumn{3}{c|}{Reddit}     & \multicolumn{3}{c|}{Amazon} & \multicolumn{3}{c}{Pubmed}  \\
                                      &                          & Acc & F1 & AUC & Acc & F1 & AUC & Acc & F1 & AUC & Acc & F1 & AUC & Acc & F1 & AUC \\ \midrule
\multirow{3}{*}{supervised}           
 & GAT                        & 84.30    & 83.35    & 85.43     & 68.63    & 82.79    & 89.98     & 93.50    & 93.03    & 94.48     & 85.00    & 82.67    & 88.78     & 80.05    & 77.07    & 79.26     \\
                                      & GCN                        & 83.85    & 84.90    & 85.90     & 66.67    & 81.01    & 89.62     & 83.50    & 84.51    & 91.43     & 89.00    & 89.81    & 98.85     & 79.00    & 77.73    & 80.19     \\
                                      & GT          & 85.95    & 86.01    & 87.25     & 69.70    & 83.03    & 82.46     & 95.50    & 94.52    & 96.89     & 94.00    & 93.62    & 99.34     & 74.50    & 65.77    & 85.19      
\\\midrule
\multirow{6}{*}{\makecell[c]{pre-train \\+\\  fine-tune}}  
& GraphCL+GAT                & 85.64    & 85.97    & 87.22     & 72.67    & 82.85    & 92.98     & 94.00    & 93.75    & 98.43     & 86.50    & 86.96    & 84.47     & 85.54    & 83.92    & 91.78     \\
                                      & GraphCL+GCN                & 86.36    & 85.82    & 86.39     & 70.67    & 81.82    & 90.00     & 94.00    & 93.94    & 97.04     & 86.50    & 84.92    & 98.41     & 80.00    & 78.05    & 85.21     \\
                                      & GraphCL+GT  & 85.79    & 86.27    & 87.51     & 86.01    & 85.38    & 88.58     & 96.67    & 95.38    & 97.65     & 96.50    & 97.42    & 98.12     & 85.50    & 87.11    & 81.68     \\
                                      & SimGRACE+GAT               & 86.85    & 86.80    & 88.12     & 85.33    & 85.26    & 90.04     & 95.50    & 95.54    & 97.11     & 87.50    & 86.34    & 88.65     & 80.01    & 81.03    & 86.89     \\
                                      & SimGRACE+GCN               & 85.62    & 85.38    & 87.83     & 89.33    & 86.34    & 95.10     & 88.00    & 87.88    & 94.49     & 98.45    & 97.57    & 98.29     & 80.50    & 82.58    & 91.22     \\
                                      & SimGRACE+GT & 86.35    & 87.03    & 88.47     & 86.00    & 89.52    & 90.42     & 97.50    & 95.54    & 96.92     & 96.50    & 96.45    & 99.09     & 81.00    & 79.57    & 85.69      
\\\midrule
\multirow{6}{*}{prompt}               
  & GraphCL+GAT                & 86.85    & 86.88    & 87.92     & 76.67    & 83.00    & 96.22     & 95.36    & 94.50    & 98.65     & 88.50    & 86.00    & 87.15     & 86.50    & 84.75    & 92.61     \\
                                      & GraphCL+GCN                & 86.87    & 86.80    & 87.79     & 76.67    & 82.37    & 93.54     & 95.50    & 95.52    & 97.75     & 86.96    & 85.63    & 98.66     & 81.50    & 78.61    & 86.11     \\
                                      & GraphCL+GT  & 87.02    & 86.90    & 87.97     & 86.67    & 88.00    & 91.10     & 97.03    & 95.94    & 98.62     & 98.50    & 98.48    & 98.53     & 86.50    & 87.78    & 82.21     \\
                                      & SimGRACE+GAT               & 87.37    & 87.33    & 88.37     & 91.33    & 92.30    & 95.18     & 95.72    & 96.69    & 97.64     & 95.50    & 95.38    & 98.89     & 80.50    & 82.03    & 87.86     \\
                                      & SimGRACE+GCN               & 86.85    & 86.80    & 88.67     & 93.47    & 97.69    & 97.08     & 88.00    & 88.12    & 95.10     & 98.50    & 98.52    & 98.55     & 81.00    & 83.76    & 91.41     \\
                                      & SimGRACE+GT & 87.30    & 87.24    & 88.74     & 95.33    & 96.52    & 94.46     & 98.00    & 98.02    & 99.38     & 98.50    & 98.52    & 99.10     & 82.50    & 80.45    & 87.61       
\\\midrule
\multicolumn{2}{c|}{IMP(\%)}                                        & 1.65     & 1.48     & 1.28      & 12.26    & 6.84     & 5.21      & 1.94     & 2.29     & 1.88      & 3.63	&3.44&	2.03  & 2.98     & 4.66     & 3.21   \\\bottomrule
\end{tabular}%
}
\end{table*}
\begin{table*}[h]
\centering
\caption{Graph-level performance (\%) with 100-shot setting. IMP (\%): the average improvement of prompt over the rest.}
\label{tab:graph_level}
\resizebox{0.99\textwidth}{!}{%
\begin{tabular}{@{}p{0.07\textwidth}<{\centering}c|p{0.025\textwidth}<{\centering}p{0.025\textwidth}<{\centering}p{0.035\textwidth}<{\centering}|p{0.025\textwidth}<{\centering}p{0.025\textwidth}<{\centering}p{0.035\textwidth}<{\centering}|p{0.025\textwidth}<{\centering}p{0.025\textwidth}<{\centering}p{0.035\textwidth}<{\centering}|p{0.025\textwidth}<{\centering}p{0.025\textwidth}<{\centering}p{0.035\textwidth}<{\centering}|p{0.025\textwidth}<{\centering}p{0.025\textwidth}<{\centering}p{0.035\textwidth}<{\centering}@{}}
\toprule
\multirow{2}{*}{\makecell[c]{Training\\ schemes}}     & \multirow{2}{*}{Methods} & \multicolumn{3}{c|}{Cora}       & \multicolumn{3}{c|}{CiteSeer}    & \multicolumn{3}{c|}{Reddit}     & \multicolumn{3}{c|}{Amazon} & \multicolumn{3}{c}{Pubmed}  \\
                                      &                          & Acc & F1 & AUC & Acc & F1 & AUC & Acc & F1 & AUC & Acc & F1 & AUC & Acc & F1 & AUC \\ \midrule
\multirow{3}{*}{supervised}      
& GAT                        & 84.40    & 86.44    & 87.60     & 86.50    & 84.75    & 91.75     & 79.50    & 79.76    & 82.11     & 93.05    & 94.04    & 93.95     & 69.86    & 72.30    & 66.92     \\
                                                                             & GCN                        & 83.95    & 86.01    & 88.64     & 85.00    & 82.56    & 93.33     & 64.00    & 70.00    & 78.60     & 91.20    & 91.27    & 94.33     & 61.30    & 59.97    & 66.29     \\
                                                                             & GT          & 85.85    & 85.90    & 89.59     & 77.50    & 75.85    & 89.72     & 69.62    & 68.01    & 66.32     & 90.33    & 91.39    & 94.39     & 60.30    & 60.88    & 67.62 
\\\midrule
\multirow{6}{*}{\makecell[c]{pre-train  \\ + \\fine-tune}}     
 & GraphCL+GAT                & 85.50    & 85.54    & 89.31     & 83.00    & 85.47    & 92.13     & 72.03    & 72.82    & 83.23     & 92.15    & 92.18    & 94.78     & 85.50    & 85.50    & 86.33     \\
                                                                             & GraphCL+GCN                & 85.50    & 85.59    & 87.94     & 86.50    & 84.57    & 94.56     & 71.00    & 71.90    & 80.33     & 93.58    & 93.55    & 94.93     & 78.75    & 77.29    & 89.40     \\
                                                                             & GraphCL+GT  & 85.95    & 85.05    & 87.92     & 84.50    & 81.87    & 88.36     & 69.63    & 70.06    & 81.35     & 91.68    & 91.55    & 94.78     & 86.85    & 86.93    & 88.91     \\
                                                                             & SimGRACE+GAT               & 86.04    & 86.33    & 88.55     & 83.50    & 85.84    & 90.09     & 81.32    & 81.64    & 88.61     & 93.58    & 93.57    & 93.91     & 87.33    & 86.70    & 88.02     \\
                                                                             & SimGRACE+GCN               & 85.95    & 86.05    & 89.33     & 84.50    & 86.46    & 91.60     & 80.50    & 81.52    & 89.11     & 90.73    & 90.52    & 94.85     & 85.26    & 84.64    & 86.99     \\
                                                                             & SimGRACE+GT & 86.40    & 86.47    & 89.64     & 81.00    & 81.54    & 89.81     & 69.50    & 70.97    & 77.11     & 92.63    & 92.56    & 94.04     & 85.95    & 86.05    & 89.37     
 \\\midrule
\multirow{6}{*}{prompt}
& GraphCL+GAT                & 86.40    & 86.47    & 89.46     & 86.50    & 89.93    & 92.24     & 73.36    & 73.32    & 84.77     & 94.08    & 94.02    & 94.20     & 85.95    & 85.97    & 87.17     \\
                                                                             & GraphCL+GCN                & 85.95    & 86.01    & 88.95     & 87.00    & 85.87    & 95.35     & 72.50    & 72.91    & 81.37     & 94.05    & 94.05    & 94.98     & 84.60    & 84.43    & 88.96     \\
                                                                             & GraphCL+GT  & 86.05    & 85.17    & 88.93     & 85.50    & 85.28    & 88.60     & 72.63    & 70.97    & 82.39     & 92.63    & 92.64    & 94.82     & 87.03    & 86.96    & 89.10     \\
                                                                             & SimGRACE+GAT               & 86.67    & 86.36    & 89.51     & 87.50    & 88.37    & 91.47     & 82.62    & 83.33    & 89.41     & 93.35    & 94.66    & 94.61     & 87.75    & 87.69    & 88.88     \\
                                                                             & SimGRACE+GCN               & 86.85    & 86.90    & 89.95     & 85.00    & 85.85    & 91.95     & 81.00    & 82.24    & 89.43     & 93.95    & 92.06    & 93.89     & 85.50    & 85.54    & 87.30     \\
                                                                             & SimGRACE+GT & 86.85    & 86.87    & 89.75     & 87.50    & 86.63    & 90.85     & 76.50    & 80.82    & 86.84     & 94.05    & 94.06    & 94.96     & 86.40    & 86.50    & 89.74      
\\\midrule
\multicolumn{2}{c|}{IMP(\%)}                                                                               & 1.12     & 0.43     & 0.79      & 3.52     & 4.54     & 0.53      & 4.69     & 4.31     & 6.13      &1.72&	1.39&	0.14 & 10.66    & 10.77    & 9.16     \\
\bottomrule
\end{tabular}%
}
\end{table*}

\end{document}